\documentstyle[11pt,publ,epsfig]{article}

\begin{document}
\vspace*{0.5cm}
\begin{flushright}
CLNS~00/1677\\[0.15cm]
hep-ph/0006265\\
June 2000
\end{flushright}
\vspace*{1.5cm}

\title{\boldmath
TOWARDS A THEORY OF HADRONIC $B$ DECAYS
\unboldmath}

\author{Matthias Neubert\footnote{Invited review talk presented at 
the XXXVth Rencontres de Moriond, {\em QCD and High-Energy Hadronic 
Interactions}, Les Arcs, France, 18--25 March 2000}}

\address{Newman Laboratory of Nuclear Studies, Cornell University\\
Ithaca, New York 14853, U.S.A.}

\maketitle\abstracts{
We review recent advances in the theory of strong-interaction effects
and final-state interactions in hadronic weak decays of heavy mesons. 
In the heavy-quark limit, the amplitudes for most nonleptonic, 
two-body $B$ decays can be calculated from first principles and 
expressed in terms of semileptonic form factors and light-cone 
distribution amplitudes. We discuss the features of this novel QCD
factorization and illustrate its phenomenological implications.}

\section{Introduction}

\begin{figure}
\centerline{\psfig{figure=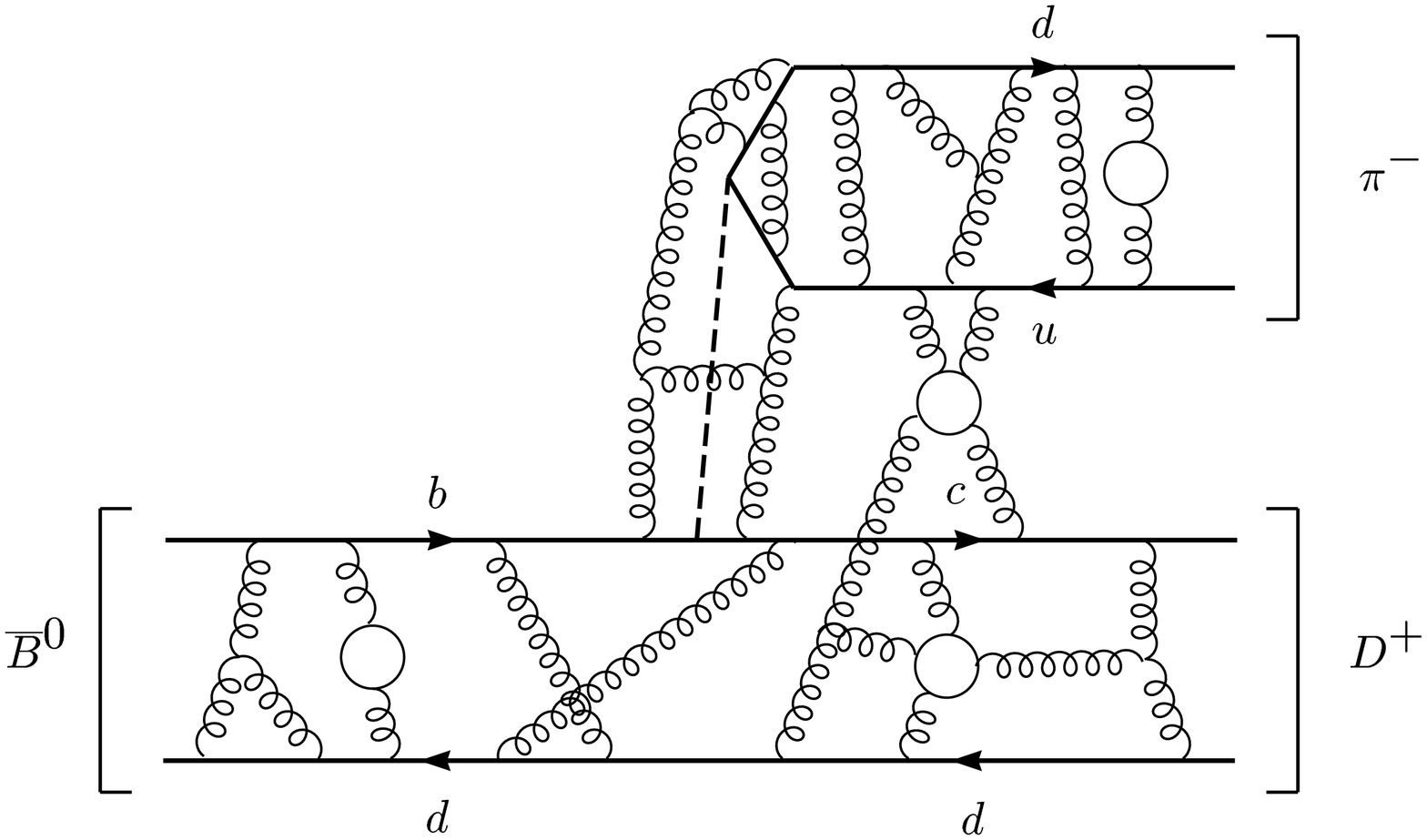,width=6.5cm}
\psfig{figure=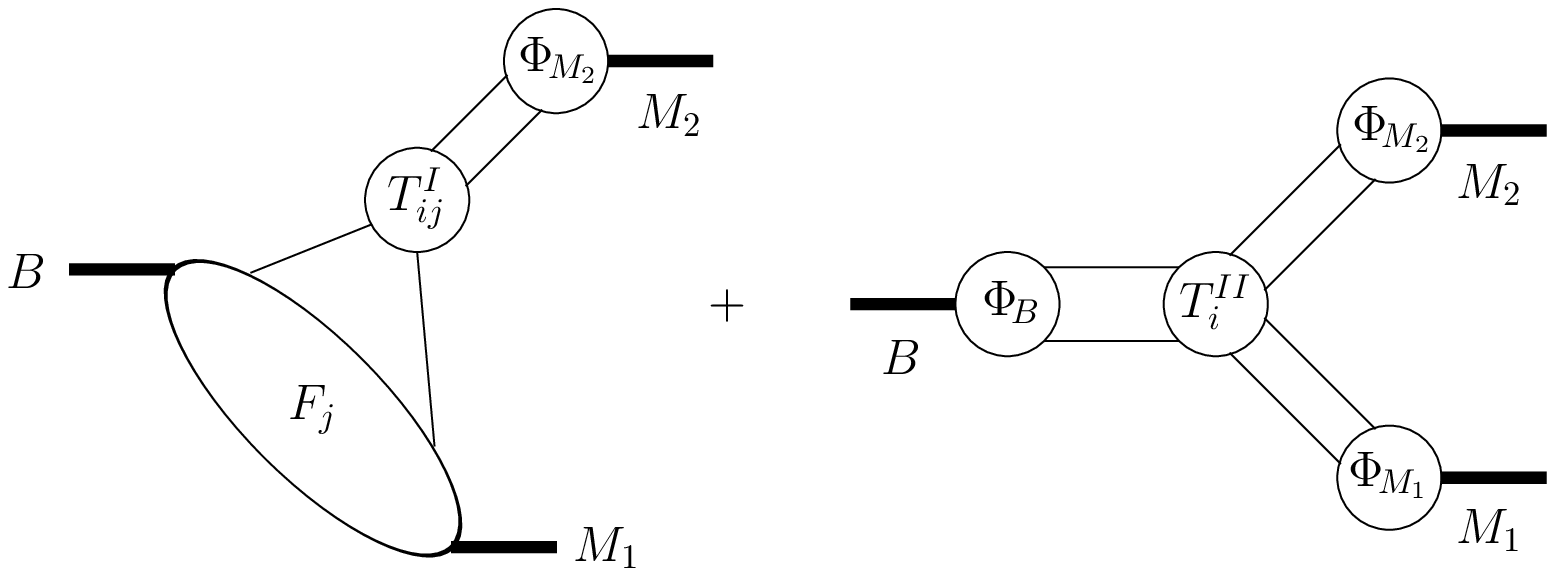,width=9.0cm}}
\caption{Left: Strong-interaction effects in a hadronic weak decay.
Right: QCD factorization in the heavy-quark limit. The second term
is power suppressed for $\bar B\to D\pi$ but must be kept for decays 
with two light mesons in the final state, such as $\bar B\to\pi\pi$.
Contributions not shown (such as weak annihilation graphs) are power
suppressed.  
\label{fig:nonlep}}
\end{figure}

The theoretical description of hadronic weak decays is difficult 
due to nonperturbative strong-interaction dynamics. This will affect
the interpretation of the data collected at the $B$ factories, 
including studies of CP violation and searches for New Physics. If 
these strong-interaction effects could be computed in a 
model-independent way, this would enhance our ability to uncover the 
origin of CP violation. The complexity of the problem is illustrated 
in the cartoon on the left-hand side of Fig.~\ref{fig:nonlep}. It is 
well known how to control the effects of hard gluons with virtuality 
between the electroweak scale $M_W$ and the scale $m_B$ characteristic 
to the decays of interest. They can be dealt with by constructing a 
low-energy effective weak Hamiltonian 
\begin{equation}\label{Heff}
   {\cal H}_{\rm eff} = \frac{G_F}{\sqrt2} \sum_i
   \lambda_i^{\rm CKM}\,C_i(M_W/\mu)\,O_i(\mu) \,,
\end{equation}
where $\lambda_i^{\rm CKM}$ are products of CKM matrix elements, 
$C_i(M_W/\mu)$ are calculable short-distance coefficients, and 
$O_i(\mu)$ are local operators renormalized at a scale 
$\mu={\cal O}(m_B)$. The challenge is to calculate the hadronic 
matrix elements of these operators with controlled theoretical 
uncertainties, using a systematic approximation scheme.

Previous field-theoretic attempts to evaluate these matrix elements 
have employed dynamical schemes such as lattice field theory, QCD sum 
rules, or the hard-scattering approach. The first two have great 
difficulties in accounting for final-state rescattering, which however 
is very important for predicting direct CP asymmetries. The 
hard-scattering approach misses the leading soft contribution to the 
$\bar B\to\mbox{meson}$ transition form factors and thus falls short 
of reproducing the correct magnitude of the decay amplitudes. In view 
of these difficulties, most previous analyses of hadronic decays have 
employed phenomenological models such as ``naive'' or ``generalized 
factorization'', in which the complicated matrix elements of four-quark 
operators in the effective weak Hamiltonian are replaced (in an {\em 
ad hoc\/} way) by products of current matrix elements. Corrections to 
this approximation are accounted for by introducing a set of 
phenomenological parameters $a_i$. A different strategy is to classify 
nonleptonic decay amplitudes according to flavor topologies (``trees'' 
and ``penguins''), which can be decomposed into SU(3) or isospin 
amplitudes. This leads to relations between different decay amplitudes 
in the flavor-symmetry limit. No attempt is made, however, to compute 
these amplitudes from first principles.

\section{QCD Factorization Formula}

Here we summarize recent progress in the theoretical understanding
of nonleptonic decay amplitudes in the heavy-quark 
limit.\cite{BBNS,bigpaper} The underlying idea is to exploit the 
presence of a large scale, i.e., the fact that 
$m_b\gg\Lambda_{\rm QCD}$. In order to disentangle the physics 
associated with these two scales, we factorize and compute hard 
contributions to the decay amplitudes arising from gluons with 
virtuality of order $m_b$, and parameterize soft and collinear 
contributions. Considering the cartoon in Fig.~\ref{fig:nonlep}, we 
denote by $M_1$ the meson that absorbs the spectator quark of the $B$ 
meson, and by $M_2$ the meson at the upper vertex, to which we refer 
as the ``emission particle''. We find that nonleptonic decay amplitudes
simplify in the heavy-quark limit if $M_2$ is a light meson. Then 
at leading power in $\Lambda_{\rm QCD}/m_b$ all long-distance 
contributions to the matrix elements can be factorized into 
semileptonic form factors and meson light-cone distribution amplitudes, 
which are much simpler quantities than the nonleptonic amplitudes 
themselves. (Light-cone distribution amplitudes enter because the 
partons in the emission particle carry large energy and are almost 
collinear.) ``Nonfactorizable'' effects connecting the partons of the 
emission particle with the rest of the diagram are dominated by hard 
gluon exchange and can be computed using perturbation theory. A 
graphical representation of the resulting ``factorization formula'' is 
shown on the right-hand side in Fig.~\ref{fig:nonlep}. The physical 
picture underlying factorization is color transparency.\cite{Bj89,DG91} 
If the emission particle is a light meson, its constituents carry large 
energy of order $m_b$ and are nearly collinear. Soft gluons coupling to 
this system see only its net zero color charge and hence decouple. 
Interactions with the color dipole of the small $q\bar q$-pair are 
power suppressed in the heavy-quark limit.

For $B$ decays into final states containing a heavy charm meson and 
a light meson $L$, the factorization formula takes the form
\begin{equation}\label{fff}
   \langle D^{(*)+} L^-|\,O_i(\mu)\,|\bar B_d\rangle
   = \sum_j F_j^{\bar B\to D^{(*)}}
   \int\limits_0^1\!\mbox{d}u\,T_{ij}^{\rm I}(u,\mu)\,\Phi_L(u,\mu)
   + {\cal O}\bigg(\frac{\Lambda_{\rm QCD}}{m_b}\bigg) \,,
\end{equation}
where $O_i$ is an operator in the effective weak Hamiltonian 
(\ref{Heff}), $F_j^{\bar B\to D^{(*)}}$ are transition form factors, 
$\Phi_L$ is the leading-twist light-cone distribution amplitude of 
the light meson, and $T_{ij}^{\rm I}$ are process-dependent 
hard-scattering kernels. For decays into final states containing two 
light mesons there is a second type of contribution to the 
factorization formula, which involves a hard interaction with the 
spectator quark in the $B$ meson. It is contained in the second 
graph on the right-hand side in Fig.~\ref{fig:nonlep}. Below we focus 
on $\bar B\to D L$ decays, where this second term is power suppressed 
and can be neglected. Decays into two light final-state mesons are 
more complicated\,\cite{BBNS} and have been discussed 
elsewhere.\cite{talks}

In order to prove factorization one must first separate hard from
infrared (soft and collinear) contributions to the decay amplitudes. 
This is done at the level of Feynman diagrams. One must then show that 
``nonfactorizable contributions'', i.e., contributions not associated 
with $\bar B\to M_1$ form factors or meson wave functions, are 
dominated by hard gluon exchange. This amounts to showing that the soft 
and collinear singularities, which are present in individual Feynman 
diagrams, cancel in the sum of all contributions. For the case of 
decays into a heavy--light final state this cancellation has been 
demonstrated by an explicit two-loop analysis.\cite{bigpaper} General 
arguments support factorization to all orders of perturbation theory. 
The fact that these cancellations occur is far from trivial, given 
that by power counting the $\bar B\to M_1$ form factors (both for a 
heavy and a light final-state meson) are dominated by soft gluon 
exchange. To complete the proof of factorization one must also show that 
contributions from transverse momenta of the partons in the final-state 
mesons, from asymmetric parton configurations where one (or several) 
partons are soft, and from non-valence Fock states (containing 
additional hard or soft partons) are power suppressed, and that 
competing flavor topologies such as weak annihilation are power 
suppressed, too. All this is discussed in detail in our recent 
work.\cite{bigpaper}

The factorization formula for nonleptonic decays provides a 
model-independent basis for the analysis of these processes in an 
expansion in powers and logarithms of $\Lambda_{\rm QCD}/m_b$. At
leading power, but to all orders in $\alpha_s$, the decay amplitudes
assume the factorized form shown in (\ref{fff}). Having such a
formalism based on power counting in $\Lambda_{\rm QCD}/m_b$ is of 
great importance to the theoretical description of hadronic weak 
decays, since it provides a well-defined limit of QCD in which these 
processes admit a rigorous, theoretical description. (For instance, 
the possibility to compute systematically $O(\alpha_s)$ corrections to 
``naive factorization'', which emerges as the leading term in the 
heavy-quark limit, solves the old problem of renormalization-scale and 
scheme dependences of nonleptonic amplitudes.) The usefulness of
this new scheme may be compared with the usefulness of the heavy-quark
effective theory for the analysis of exclusive semileptonic decays of
heavy mesons, or of the heavy-quark expansion for the analysis of
inclusive decay rates. In all three cases, it is the fact that 
hadronic uncertainties can be eliminated up to power corrections
in $\Lambda_{\rm QCD}/m_b$ that has advanced our ability to control 
theoretical errors.

However, it must be stressed that we are just beginning to explore
the theory of nonleptonic $B$ decays. Some important conceptual 
problems remain to be better understood. In the next few years it 
will be important to further develop this novel approach. This should 
include proving factorization at leading power to all orders in 
$\alpha_s$, developing a formalism for dealing with power corrections 
to factorization, understanding the light-cone structure of heavy 
mesons, and understanding the relevance (or irrelevance) of Sudakov 
form factors.\cite{Li} Also, we must gauge the accuracy of the approach 
by learning about the magnitude of corrections to the heavy-quark limit
from extensive comparisons of theoretical predictions with data. As
experience with previous heavy-quark expansions has shown, this is 
going to be a long route. Yet, already we have obtained important 
insights. Let us mention three points here:

1. Corrections to ``naive factorization'' (usually called 
``nonfactorizable effects'') are process dependent, in contrast with 
a basic assumption underlying models of ``generalized factorization''. 

2. The physics of nonleptonic decays is both rich and complicated. In
general, it is characterized by an interplay of several small parameters 
(Wilson coefficients, CKM factors, $1/N_c$, etc.) in addition to the 
small parameter $\Lambda_{\rm QCD}/m_b$ relevant to QCD factorization. 
In some cases, terms that are formally power suppressed may be enhanced 
by factors such as $2m_\pi^2/(m_u+m_d)\approx 3$\,GeV, which are larger 
than naive dimensional analysis would suggest.\cite{BBNS} Finally, 
several not-so-well-known input parameters (e.g., heavy-to-light form 
factors and light-cone distribution amplitudes) introduce sizable 
numerical uncertainties in the predictions.

3. Strong-interaction phases arising from final-state interactions are
suppressed in the heavy-quark limit. More precisely, the imaginary 
parts of nonleptonic decay amplitudes are suppressed by at least one 
power of $\alpha_s(m_b)$ or $\Lambda_{\rm QCD}/m_b$. At leading power, 
the phases are calculable from the imaginary parts of the 
hard-scattering kernels in the factorization formula. Since this 
observation is of paramount importance to the phenomenology of
direct CP violation, we will discuss it in some more detail.

\section{Final-State Interactions and Rescattering Phases}

Final-state interactions are usually discussed in terms of intermediate 
hadronic states. This is suggested by the unitarity relation (taking 
$\bar B\to\pi\pi$ for definiteness) 
\begin{equation}\label{unitarity}
   \mbox{Im}\,{\cal A}_{\bar B\to\pi\pi} \sim \sum_n 
   {\cal A}_{\bar B\to n}\,{\cal A}_{n\to\pi\pi}^* \,.
\end{equation}
However, because of the dominance of hard rescattering in the 
heavy-quark limit we can also interpret the sum as extending over 
intermediate states of partons. In the limit $m_b\to\infty$ the number 
of physical intermediate states is arbitrarily large. We may then 
argue on the grounds of parton--hadron duality that their average is 
described well enough (up to $\Lambda_{\rm QCD}/m_b$ corrections, say) 
by a partonic calculation. This is the picture implied by the 
factorization formula. The hadronic language is in principle exact. 
However, the large number of intermediate states makes it intractable 
to observe systematic cancellations, which usually occur in an 
inclusive sum over hadronic states. A example familiar from previous 
applications of the heavy-quark expansion is the calculation of the 
inclusive semileptonic decay width of a heavy hadron. Here the leading 
term is given by the free quark decay, but attempts to reproduce this 
obvious result by summing over exclusive modes has been successful only 
in two-dimensional toy models,\cite{GL97,Bigi99} not in QCD.

In many phenomenological discussions of final-state interactions it 
has been assumed that systematic cancellations are absent. It is then 
reasonable to consider the size of rescattering effects for a subset of 
intermediate states (such as the two-body states), assuming that this 
will provide a correct order-of-magnitude estimate for the total 
rescattering effect. This strategy underlies all estimates of 
final-state phases using dispersion relations and Regge 
phenomenology.\cite{DGPS96,Fa97} These approaches suggest that soft 
rescattering phases do not vanish in the heavy-quark limit. However, 
they also leave open the possibility of systematic cancellations.
The QCD factorization formula implies that systematic cancellations 
do indeed occur in the sum over all intermediate states. The underlying 
physical reason is that the sum over all states is accurately 
represented by a $q\bar{q}$ fluctuation in the emitted light meson 
of small transverse size of order $1/m_b$. Because the $q\bar{q}$ pair 
is small, the physics of rescattering is very different from elastic 
$\pi\pi$ scattering, and hence the Regge phenomenology applied to $B$ 
decays is difficult to justify in the heavy-quark limit. Consequently, 
the numerical estimates for rescattering effects and final-state phases 
obtained using Regge models are likely to overestimate the correct size 
of the effects.

\section{\boldmath
Applications to $\bar B_d\to D^{(*)+} L^-$ Decays
\unboldmath}

Our results for the nonleptonic $\bar B_d\to D^{(*)+}L^-$ decay 
amplitudes (with $L$ a light meson) can be compactly expressed in 
terms of the matrix elements of a ``transition operator''
\begin{equation}\label{heffa1}
   {\cal T} = \frac{G_F}{\sqrt2}\,V^*_{ud} V_{cb}
   \left[ a_1(D L)\,Q_V - a_1(D^* L)\,Q_A \right] \,,
\end{equation}
where the hadronic matrix elements of the operators 
$Q_V=\bar c\gamma^\mu b\,\otimes\,\bar d\gamma_\mu(1-\gamma_5)u$
and $Q_A=\bar c\gamma^\mu\gamma_5 b\,\otimes\,
\bar d\gamma_\mu(1-\gamma_5)u$ are understood to be evaluated in 
factorized form. Eq.~(\ref{heffa1}) defines the quantities 
$a_1(D^{(*)} L)$, which include the leading ``nonfactorizable'' 
corrections, in a renormalization-group invariant way. To leading power 
in $\Lambda_{\rm QCD}/m_b$ these quantities should not be interpreted as 
phenomenological parameters (as is usually done), because they are 
dominated by hard gluon exchange and thus calculable in QCD. At 
next-to-leading order in $\alpha_s$, we obtain\,\cite{bigpaper}
\begin{equation}\label{a1}
   a_1(D^{(*)}L) = \bar C_1(m_b) + \frac{\bar C_2(m_b)}{N_c} \left[
   1 + \frac{C_F\alpha_s(m_b)}{4\pi} \int\limits^1_0\!\mbox{d}u\,
   F(u,\pm z)\,\Phi_L(u) \right] \,,
\end{equation}
where $\bar C_i(m_b)$ are the so-called ``renormalization-scheme 
independent'' Wilson coefficients, $z=m_c/m_b$, and the upper (lower) 
sign in the second argument of the function $F(u,\pm z)$ refers to a 
$D$ ($D^*$) meson in the final state. An exact analytic expression for 
this function is known but not relevant to our discussion here. Note 
that the coefficients $a_1(D L)$ and $a_1(D^* L)$ are nonuniversal, 
i.e., they are explicitly dependent on the nature of the final-state 
mesons. Politzer and Wise have computed the ``nonfactorizable'' vertex 
corrections to the ratio of the $\bar B_d\to D^+\pi^-$ and 
$\bar B_d\to D^{*+}\pi^-$ decay rates.\cite{PW91} This requires the 
symmetric part (with respect to $u\leftrightarrow 1-u$) of the 
difference $F(u,z)-F(u,-z)$. We agree with their result.

The expressions for the decay amplitudes obtained by evaluating the 
hadronic matrix elements of the transition operator ${\cal T}$ involve
products of CKM matrix elements, light-meson decay constants, 
$\bar B\to D^{(*)}$ transition form factors, and the QCD parameters 
$a_1(D^{(*)} L)$. A numerical analysis shows that 
$|a_1|=1.055\pm 0.025$ for the decays considered below. Below we will 
use this as our central value.

\subsection{Test of factorization}

A particularly clean test of our predictions is obtained by relating 
the $\bar B_d\to D^{*+} L^-$ decay rates to the differential 
semileptonic $\bar B_d\to D^{*+}\,l^-\nu$ decay rate evaluated at 
$q^2=m_L^2$. In this way the parameters $|a_1(D^* L)|$ can be measured 
directly.\cite{Bj89} One obtains
\begin{equation}
   \frac{\Gamma(\bar B_d\to D^{*+} L^-)}
    {d\Gamma(\bar B_d\to D^{*+} l^-\bar\nu)/dq^2\big|_{q^2=m^2_L}}
   = 6\pi^2 |V_{ud}|^2 f^2_L\,|a_1(D^* L)|^2 \,.
\end{equation}
With our result for $a_1$ this relation becomes a prediction based on 
first principles of QCD. This is to be contrasted with the usual 
interpretation of this formula, where $a_1$ plays the role of a 
phenomenological parameter that is fitted from data.

Using data reported by the CLEO Collaboration,\cite{Rodr97} we find
\begin{equation}
   |a_1(D^*\pi)| = 1.08 \pm 0.07 \,, \quad
   |a_1(D^*\rho)| = 1.09\pm 0.10 \,, \quad
   |a_1(D^* a_1)| = 1.08\pm 0.11 \,,
\end{equation}
in good agreement with our prediction. It is reassuring that the data 
show no evidence for large power corrections to our results. However, 
a further improvement in the experimental accuracy would be desirable 
in order to become sensitive to process-dependent, nonfactorizable 
effects.

\subsection{Predictions for class-I decay amplitudes}

We now consider a larger set of so-called class-I decays of the form 
$\bar B_d\to D^{(*)+} L^-$, all of which are governed by the transition 
operator (\ref{heffa1}). In Table~\ref{tab:10decays} we compare
the QCD factorization predictions with experimental data. As previously 
we work in the heavy-quark limit, i.e., our predictions are model 
independent up to corrections suppressed by at least one power of 
$\Lambda_{\rm QCD}/m_b$. There is good agreement between the predictions 
and the data within experimental errors, which however are still large. 
It would be desirable to reduce these errors to the percent level. Note 
that we have not attempted to adjust the semileptonic form factors 
$F_+^{\bar B\to D}$ and $A_0^{\bar B\to D^*}$ entering our results so 
as to obtain a best fit to the data. (The fact that with 
$F_+(0)=A_0(0)=0.6$ our predictions for the $\bar B_d\to D^{(*)+}\pi^-$ 
branching ratios come out higher than the central experimental results 
reported by the CLEO Collaboration must not be taken as evidence 
against QCD factorization. As we have seen above, the value of 
$|a_1(D^*\pi)|$ extracted in a form-factor independent way is in good 
agreement with our theoretical result.)

\begin{table}
\caption{\label{tab:10decays}
Model-independent predictions for the branching ratios (in units of
$10^{-3}$) of $\bar B_d\to D^{(*)+} L^-$ decays in the heavy-quark 
limit. Predictions are in units of $(|V_{cb}|/0.04)^2\times
(|a_1|/1.05)^2\times(\tau_{B_d}/1.56\,\mbox{ps})$. We show experimental 
results reported by the CLEO Collaboration~\protect\cite{CLEO9701} and 
the Particle Data Group.\protect\cite{PDG}}
\vspace{0.4cm}
\begin{center}
\begin{tabular}{|l|c|cc|}
\hline
&&&\\[-0.4cm]
Decay mode & Theory (HQL) & CLEO data & PDG98~ \\
&&&\\[-0.4cm]
\hline
&&&\\[-0.4cm]
$\bar B_d\to D^+\pi^-$   & $3.27\times[F_+(0)/0.6]^2$ & $2.50\pm 0.40$
 & $3.0\pm 0.4$ \\
$\bar B_d\to D^+ K^-$    & $0.25\times[F_+(0)/0.6]^2$ & --- & --- \\
$\bar B_d\to D^+\rho^-$  & $7.64\times[F_+(0)/0.6]^2$ & $7.89\pm 1.39$
 & $7.9\pm 1.4$ \\
$\bar B_d\to D^+ K^{*-}$ & $0.39\times[F_+(0)/0.6]^2$ & --- & --- \\
$\bar B_d\to D^+ a_1^-$  & $7.76\times[F_+(0)/0.6]^2$ & $8.34\pm 1.66$
 & $6.0\pm 3.3$ \\
&&&\\[-0.4cm]
\hline
&&&\\[-0.4cm]
$\bar B_d\to D^{*+}\pi^-$   & $3.05\times[A_0(0)/0.6]^2$
 & $2.34\pm 0.32$ & $2.8\pm 0.2$ \\
$\bar B_d\to D^{*+} K^-$    & $0.22\times[A_0(0)/0.6]^2$ & --- & --- \\
$\bar B_d\to D^{*+}\rho^-$  & $7.59\times[A_0(0)/0.6]^2$
 & $7.34\pm 1.00$ & $6.7\pm 3.3$ \\
$\bar B_d\to D^{*+} K^{*-}$ & $0.40\times[A_0(0)/0.6]^2$ & --- & --- \\
$\bar B_d\to D^{*+} a_1^-$  & $8.53\times[A_0(0)/0.6]^2$
 & $11.57\pm 2.02$ & $13.0\pm 2.7$ \\[0.1cm]
\hline 
\end{tabular}
\end{center}
\end{table}

The observation that the experimental data on class-I decays into 
heavy--light final states show good agreement with our predictions may 
be taken as (circumstantial) evidence that in these decays there are no 
unexpectedly large power corrections. In our recent work\,\cite{bigpaper} 
we have addressed the important question of power corrections 
theoretically by providing estimates for two sources of power-suppressed 
effects: weak annihilation and nonfactorizable spectator interactions. 
We stress that a complete account of power corrections to the 
heavy-quark limit cannot be performed in a systematic way, since these 
effects are no longer dominated by hard gluon exchange. However, we 
believe that our estimates are nevertheless instructive. 

We parameterize the annihilation contribution to the 
$\bar B_d\to D^+\pi^-$ decay amplitude in terms of an amplitude $A$ 
such that ${\cal A}(\bar B_d\to D^+\pi^-)=T+A$, where $T$ is the ``tree 
topology'', which contains the dominant factorizable contribution. We 
find that $A/T\sim 0.04$, indicating that the annihilation contribution 
is a correction of a few percent. We have also obtained an estimate of 
nonfactorizable spectator interactions, which are part of the $T$, 
finding that $T_{\rm spec}/T_{\rm lead}\sim-0.03$. In both cases, the 
results exhibit the expected linear power suppression 
$\sim\Lambda_{\rm QCD}/m_b$ in the heavy-quark limit. We conclude that 
the typical size of power corrections to the heavy-quark limit in 
class-I $B$ decays into heavy--light final states is at the level of 
10\% or less, and thus our predictions for the values and the near 
universality of the parameters $a_1$ governing these decay modes appear 
robust. 

\subsection{Remarks on class-II and class-III decay amplitudes}

In the class-I decays $\bar B_d\to D^{(*)+} L^-$ considered above, 
the flavor quantum numbers of the final-state mesons ensure that only
the light meson $L$ can be produced by the $(\bar d u)$ current 
contained in the operators of the effective weak Hamiltonian 
(\ref{Heff}). The factorization formula then predicts that the
corresponding decay amplitudes are factorizable in the heavy-quark 
limit. It also predicts that other topologies, in which the heavy 
charm meson is created by a $(\bar c u)$ current, are power 
suppressed. To study these topologies one may consider decays with a 
neutral charm meson in the final state. In the class-II decays 
$\bar B_d\to D^{(*)0} L^0$ the only possibility is to have the 
charm meson as the emission particle, whereas for the class-III 
decays $B^-\to D^{(*)0} L^-$ both final-state mesons can be the 
emission particle. The factorization formula predicts that in the 
heavy-quark limit class-II decay amplitudes are power suppressed 
with respect to the corresponding class-I amplitudes, whereas class-III
amplitudes should be equal to the corresponding class-I amplitudes
up to power corrections. 

Experimental data indicate sizable corrections to these predictions, 
which are mainly due to significant heavy-quark scaling violations in 
the values of the semileptonic form factors and meson decay constants. 
For a detailed analysis of this problem the reader is referred to our 
recent work.\cite{bigpaper} Note that, whereas the QCD factorization 
formula (\ref{fff}) allows us to compute the coefficients $a_1$ in the 
heavy-quark limit, it does not allow us to compute the corresponding 
parameters $a_2$ in class-II decays. Because in these decays the 
emission particle is a heavy charm meson, the mechanism of color 
transparency is not operative. For a rough estimate of $a_2$ in 
$\bar B\to\pi D$ decays we have considered the limit in which the charm 
meson is treated as a light meson, however with a highly asymmetric 
distribution amplitude. In this limit we found that 
$a_2 \approx 0.25\,e^{-i 41^\circ}$ with large theoretical 
uncertainties.\cite{bigpaper} Remarkably, this crude estimate indicates 
a significant correction to naive factorization (which gives 
$a_2\approx 0.12$). It yields the right order of magnitude for $|a_2|$ 
and, at the same time, a large strong-interaction phase.

\section{Summary and Outlook}

With the recent commissioning of the $B$ factories and the planned 
emphasis on heavy-flavor physics in future collider experiments, the 
role of $B$ decays in providing fundamental tests of the Standard Model 
and potential signatures of New Physics will continue to grow. In many 
cases the principal source of systematic uncertainty is a theoretical 
one, namely our inability to quantify the nonperturbative QCD effects 
present in these decays. This is true, in particular, for almost all 
measurements of direct CP violation. Our work provides a rigorous 
framework for the evaluation of strong-interaction effects for a large 
class of exclusive, two-body nonleptonic decays of $B$ mesons. It 
gives a well-founded field-theoretic basis for phenomenological studies 
of exclusive hadronic $B$ decays and a formal justification for the 
ideas of factorization.

We hope that the factorization formula (\ref{fff}) and its 
generalization to decays into two light mesons will form the basis
for future phenomenological studies of nonleptonic $B$ decays. We 
stress, however, that a considerable amount of conceptual work remains 
to be completed. Theoretical investigations along the lines discussed
here should be pursued with vigor. We are confident that, ultimately, 
this research will result in a {\em theory\/} of nonleptonic $B$ decays, 
which should be as useful for this area of heavy-flavor physics as the 
large-$m_b$ limit and the heavy-quark effective theory were for the 
phenomenology of semileptonic weak decays.

\section*{Acknowledgments}
It is a pleasure to thank M.~Beneke, G.~Buchalla and C.~Sachrajda for 
an ongoing collaboration on the subject of this talk. This work was 
supported in part by the National Science Foundation.

\end{document}